\documentclass[prd,aps,showpacs,tightenlines,notitlepage,nofootinbib,superscriptaddress,groupedaddress]{revtex4-1}
\usepackage{mathrsfs}
\usepackage{amsmath}
\usepackage{amsfonts}
\usepackage{array}
\usepackage{verbatim}
\usepackage{epsfig}
\usepackage{graphicx} 
\usepackage[usenames, dvipsnames]{color}
\usepackage{marginnote}
\usepackage{slashed}
\usepackage{bm}

\newcounter{mynote}

\renewcommand{\vec}[1]{\ensuremath{\boldsymbol{#1}}}

\newcommand{\bra}[1]{\ensuremath{\left< #1\,\right|}}
\newcommand{\ket}[1]{\ensuremath{\left|\, #1\right>}}

\begin{document}
\title{
Classicalization and unitarization of wee partons in QCD and Gravity:\\ The CGC-Black Hole correspondence}
\author{Gia Dvali$^{1,2}$}
\author{Raju Venugopalan$^3$}
\affiliation{$^1$Arnold Sommerfeld Center, Ludwig-Maximilians-Universit\"{a}t, Theresienstrasse 37, 80333 M\"{u}nchen, Germany\\
$^2$Max-Planck-Institut f\"{u}r Physik, F\"{o}hringer Ring 6, 80805 M\"{u}nchen, Germany\\
$^3$Physics Department, Brookhaven National Laboratory,
Bldg. 510A, Upton, NY 11973, U.S.A.
}

\begin{abstract}
 We discuss a remarkable correspondence between the description of Black Holes as highly occupied condensates of $N$ weakly interacting gravitons and that of  Color Glass Condensates (CGCs)  as highly occupied gluon states. In both cases, the dynamics of ``wee partons" in Regge asymptotics is controlled by emergent semi-hard scales that lead to perturbative unitarization and classicalization of $2\rightarrow N$ particle amplitudes at weak coupling.  In particular, they 
attain a maximal entropy permitted by unitarity, bounded by the inverse coupling $\alpha$ of the respective constituents.  
 Strikingly, this entropy is equal to the area measured in units of the Goldstone constant corresponding to the spontaneous breaking of Poincar{\'{e}} symmetry by the 
corresponding graviton or gluon condensate.  In gravity, the Goldstone constant is the Planck scale, and gives rise to the Bekenstein-Hawking entropy.   
 Likewise, in the CGC, the corresponding Goldstone scale is determined by the onset of gluon screening.
We point to further similarities in Black Hole formation, thermalization and decay, to that of the Glasma matter formed from colliding CGCs in ultrarelativistic nuclear collisions, which decays into a Quark-Gluon Plasma. 
\end{abstract}

\maketitle

\section{Introduction}

A topic of fundamental interest in physics is the possible existence of deep connections between the infrared behavior of the strong interaction and gravity, going back to the formulations of QCD and string theory. In the last couple of decades, the paradigmatic example has been the AdS/CFT correspondence, albeit both the AdS and the CFT sides of this correspondence are not the theories of phenomenological interest~\cite{Maldacena:1997re}. A striking recent example of a concrete connection between gravity and QCD is the so-called BCJ ``double copy" whereby perturbative gravity amplitudes are constructed out of QCD amplitudes and additional kinematic factors~\cite{Bern:2008qj}. 

In this work, we will discuss a possible novel  correspondence between the two theories respectively arising from our observations of remarkable similarities in the weakly coupled graviton and gluon (``parton") states of both theories at an infrared fixed point characterized by high occupancy. In the case of gravity, such high occupancy states provide a microscopic ``quantum $N$-portrait"  of a Black Hole (BHNP)~\cite{Dvali:2011aa, Dvali:2012en}. In QCD,  the corresponding quantum portrait of the wavefunction of a hadron at high energies is a saturated gluon state commonly called a Color Glass Condensate (CGC)~\cite{Gelis:2010nm}. In the collisions of two nuclei at ultrarelativistic energies, the collision of the two CGC gluon shock waves generates an overoccuped Glasma nonequilibrium state~\cite{Lappi:2006fp,Gelis:2006dv} which subsequently evolves into a Quark-Gluon Plasma (QGP)~\cite{Rischke:2003mt}.  

As we shall elaborate, the similarities between the two quantum portraits strongly hint at universal behavior that may be independent of the details of their microscopic dynamics or the initial conditions; the physics of these objects is defined by the fact that they represent {\it saturated} states corresponding to highly occupied soft quanta 
at a ``critical packing" sufficient to unitarize the cross-section for their formation.  The critical packing of the quanta (gravitons or gluons) is defined by a characteristic saturation momentum scale, denoted by $Q_S$, with the associated de Broglie wavelength $R_S = 1/Q_S$.  Other key quantities are, {\it 1)}  The occupation number 
of quanta per de Broglie volume ($R_S^3 =Q_S^{-3}$), to be denoted
by $N$,  and {\it 2)} the running coupling constant $\alpha(Q_S)$, determining the strength of the interaction amongst these quanta. 
 Critical packing is reached when the occupation number and the coupling satisfy,
\begin{equation}\label{NA}
 N = \frac{1}{\alpha(Q_S)}\,. 
 \end{equation}
As suggested by this expression, the running of  $\alpha(Q_S)$ is evaluated 
at the scale $Q_S$. The physical meaning of  critical packing 
is that it represents the point of optimal balance between the kinetic 
energies of the individual constituents and their potential  energies. In particular, in the BHNP, this is the regime in which the graviton condensate 
forms a self-sustained long-lived bound state, a Black Hole.
  
In both systems (BHNP or CGC), the critical packing results in the emergence of a 
large number of gapless modes corresponding to a specific micro-state entropy  $S$. 
An important ingredient in the correspondence between the systems is the 
relation between this micro-state entropy and the coupling $\alpha(Q_S)$.  
This relation, proposed and worked out in a series of papers 
\cite{Dvali:2019jjw, Dvali:2019ulr, Dvali:2020wqi}  constitutes an 
upper bound imposed by unitarity on the entropy: 
 \begin{equation}\label{USB}
 S_{\rm max}  =  \frac{1}{\alpha(Q_S)}\,. 
 \end{equation}  
According to \cite{Dvali:2020wqi}, the correlation between the saturation of unitarity and the bound in Eq.~(\ref{USB}) 
 is non-perturbative in nature and cannot be avoided by 
 resummations in perturbation theory. 
It was argued further that any system saturating this entropy bound shares certain universal properties with a Black Hole.  
Perhaps the most striking is the relation between the entropy 
and surface area and the role of a Goldstone boson in this relation. 
The bottom line is 
that  the maximal entropy permitted by unitarity
is equal to the area  ($\sim R_S^2$) of the system 
measured in units of a Goldstone decay constant $f$~\cite{Dvali:2020wqi,Dvali:2019jjw,Dvali:2019ulr},
 \begin{equation}\label{Gf}
 S_{\rm max}  =  {\rm  Area} \times f^2 \,. 
 \end{equation}
This Goldstone boson is a universal consequence of the spontaneous breaking of Poincar{\'{e}} symmetry. 
 For a saturated system, the expression in  Eq.~(\ref{Gf}) is equivalent to that in Eq.~(\ref{USB}).  
 To summarize Ref.~\cite{Dvali:2020wqi}, the information capacity of a unitary system is bounded by the area measured in units of the Goldstone scale. 
 
In this respect,
Black Holes are not special and their entropy 
exhibits the area-law of a generic saturated system. In the references noted, the connection 
 between  Black Holes and similar saturated objects in other physical systems was demonstrated for a number of systems,  including states in gauge theories with large number of colors. 
 Note that in the Black Hole case the Goldstone is a graviton;  the expression in Eq.~(\ref{Gf}) 
gives the Bekenstein-Hawking entropy~\cite{Bekenstein:1973ur, Hawking:1976de}. 
Other analogous universal features of such states include decay time scales and that of information recovery \cite{Dvali:2020wqi}, 
which shall be discussed later.   

A unique, and arguably the most striking, feature of the correspondence we wish to highlight here is that both the CGC state probed in QCD at high energies and the BHNP description of Black Holes occur in physical systems in nature.  The commonality in critical packing, and as we shall discuss in the relations in Eqs.~(\ref{USB}) and(\ref{Gf}), determines the similarities in important features of the two systems but the 
 connections extend beyond and cover several other key features of their dynamics.   Both Black Hole and CGC/Glasma states are consequences of the ``classicalization" and unitarization of $2\rightarrow N$ amplitudes in gravity and QCD respectively in the high energy Regge limit where the naive $N!/N^N\sim \exp(-N)$ suppression of  cross-sections is compensated by an $\exp(N)$ factor from counting the number of micro-states in the system. In the case of Black Holes, the logarithm of this number is the Black Hole entropy factor. In the CGC scattering picture, it is commonly understood to arise from the exponentiation of soft gluon bremsstrahlung. 

The idea of the saturation of trans-Planckian amplitudes 
by Black Hole production goes back to the papers by 't Hooft \cite{tHooft:1984kcu,tHooft:1996rdg}, Gross and Mende \cite{Gross:1987ar,Gross:1987kza} and 
by Amati, Ciafaloni and Veneziano  \cite{Amati:1987uf}, followed by a large body of literature we shall not attempt to review here. The step that is crucial for our
CGC/BHNP correspondence is the realization \cite{Dvali:2011aa} that the relevant  amplitude describing Black Hole formation in the scattering
of two trans-Planckian gravitons is that of the $2\rightarrow N$ Regge scattering process in which the bulk of the center of mass energy, $\sqrt{s}$, 
 is redistributed among $N$ quanta of roughly equal softness, corresponding to momenta 
 $|\vec{p}| \sim \frac{\sqrt{s}}{N}$ \cite{Dvali:2014ila}.  A two-dimensional effective field theory for $2\rightarrow N$ scattering amplitudes in this regime of ``multi-Regge" kinematics in QCD and gravity was developed by Lipatov~\cite{Lipatov:1991nf}, who also observed early-on that gravitational effective vertices could be represented as ``squares" of so-called Lipatov vertices in QCD~\cite{Lipatov:1982vv}.
 
Explicit 
computations \cite{Dvali:2014ila}
of this process both in string theory as well as in field theory 
show that when the momenta of final-state gravitons 
are given by 
$|\vec{p}| \sim  {M_P^2 \over \sqrt{s}}$,
the scattering rate 
delivers precisely the right exponential suppression factor
${\rm e}^{-N}$  that compensates the micro-state degeneracy factor ${\rm e}^{N}$ of a Black Hole of mass $M_{\rm BH} =\sqrt{s}$.
Note  that this typical momentum is equal to the inverse Schwarzschild radius of such a  Black Hole,  $|\vec{p}| = \frac{1}{R_S}$. 
This momentum also corresponds to the energy of Hawking quanta (with temperature $T_H\sim 1/R_S$) produced during its 
evaporation.  At the same time, as noted, $N$ is equal to Black Hole entropy. 
Thus without any additional input, the saturation of the $2\rightarrow N$ scattering amplitude in the Regge limit delivers key ingredients of the Black Hole $N$-portrait.  
  
 This result was further refined in an explicit computation of $2\rightarrow N$ reggeized graviton amplitudes by Addazi, Bianchi and Veneziano~\cite{Addazi:2016ksu}. 
 Perturbative unitarization in this computation is imposed through the eikonalization of the reggeon amplitudes at the unitarization boundary $R_S$ or equivalently at $T_H\sim 1/R_S$. An essential feature in demonstrating the equivalence of this picture to the quantum $N$-portrait \cite{Dvali:2011aa} is the exponentiation of real and virtual emissions of soft gravitons in the computation of \cite{Addazi:2016ksu}.

In QCD, the cross-section for $2\rightarrow N$ scattering in multi-Regge kinematics is given by the BFKL equation~\cite{Kuraev:1977fs,Balitsky:1978ic}.  As shown by Lipatov and collaborators, infrared $t$- channel divergences cancel for this ``BFKL Pomeron" in multi-Regge kinematics,  giving a result for the cross-section that corresponds to a resummation of 
leading logarithms $\alpha_S\ln(x)$, where $x$ denotes the longitudinal momentum fraction carried by wee partons in multi-Regge kinematics\footnote{In the Regge limit, $s\rightarrow \infty$, $s/t\rightarrow -\infty$, where $-t = {\rm fixed}$ is the momentum transfer squared in the gluon scattering amplitude; multi-Regge denotes Regge kinematics applied to each rung of the $2\rightarrow N$ ladder corresponding to bremsstrahlung emission. Equivalently, in the language of Deeply Inelastic Scattering (DIS), the Regge limit corresponds to Bjorken $x_{\rm Bj} \approx Q^2/s \rightarrow 0$, where $Q^2$ is the ultraviolet resolution scale of the DIS probe. In the parton model, $x_{\rm Bj}\approx x$, where $x$ is the light cone momentum fraction of a hadron carried by a parton. }. This resummation leads to a cross-section that grows rapidly with the center-of-mass energy. However when gluon occupancies $N\gg 1$, many-body processes corresponding to the screening and recombination of small $x$ gluons become important 
 and 
counter soft gluon bremsstrahlung, a phenomenon known as gluon saturation~\cite{Gribov:1984tu,Mueller:1985wy}. 

Gluon saturation leads to the classicalization of $2\rightarrow N$ amplitudes~\cite{McLerran:1993ni,McLerran:1993ka} in the Regge limit of QCD and the wee gluon matter characterizing this saturated state is the CGC. A key feature of the CGC is that the typical momenta of wee gluons resolved by a probe with resolution $Q$ is given by $Q\equiv Q_S(x)$, where $Q_S(x)$ is an emergent energy-dependent scale whose inverse, given by $R_S\sim 1/Q_S$, denotes the close packing radius of saturated gluons at maximal occupancy.  Our choice of $R_S$ for this scale is deliberate since, as we will discuss shortly, the dynamics of wee partons in both QCD and gravity is governed by this scale. We note that parton saturation and the saturation scale were not considered in early discussions of the similarities between wee parton dynamics in QCD and gravity~\cite{Verlinde:1991iu,Verlinde:1993te,Susskind:1994vu}. Taking this physics into account qualitatively modifies that discussion.

This paper is organized as follows. In section II, we will begin by noting the common features of saturated wee parton states in QCD. We will also discuss a fundamental difference in 
the UV (ultraviolet) completeness of the two theories, and discuss the sense in which the UV regime of gravity is actually probing the deep infrared (IR) regime of the theory. We will then establish the concrete dictionary of the BHNP-CGC correspondence. With this in hand, in Section III we discuss how the entropy bounds 
Eq.~(\ref{USB}) and Eq.~(\ref{Gf}) 
are saturated at the unitarization boundary both for Black Holes and the CGC. In particular, we discuss how this entropy can be expressed as an area law in units of the Goldstone constant characterizing the spontaneous breaking of Poincar{\'{e}} invariance by both saturated states. 
In Section IV, we discuss the timescales characterizing the decay of Black Holes in the BHNP and relate these to those in the CGC/Glasma; the relevant scattering rates are controlled by the dynamical screening of scattering between the wee partons mediated by collective modes of the $N$ particle state. While the thermalization process is analogous in the semi-classical evolution of the two systems, their evolution differs qualitatively after a quantum break time when the occupancies in the two systems are no longer large. We also observe similarities in the very early-stage behavior of the system whereby the rapid scrambling of information leads to a quasi-particle picture with the subsequent generation of entanglement generated by kinetic scattering processes. We end with an outlook on further aspects of the BHNP-CGC correspondence and outline future directions of work to establish this correspondence on a quantitative footing. The appendix outlines a general argument relating the Goldstone decay constant of collective modes to the area-law entropy in the two systems. 
 \section{Dictionary between the semi-classical portrait of a Black Hole and a CGC}  

We will establish here a dictionary between the semi-classical UV/IR portraits of Black Holes and CGCs, by first identifying the correspondence in the emergent scales 
 of the two theories. 
 \subsection{Black Holes} 
According to the Black Hole $N$-portrait (BHNP) \cite{Dvali:2011aa,
Dvali:2012en}, 
the Black Hole represents a long lived state of 
soft gravitons with very high occupation number $N \gg 1$; it can therefore be viewed as a graviton condensate that is approximately classical. 
For any given $N$, the graviton  wavelength $R_S$ (and thus momentum $Q_S=  R_S^{-1}$) is determined by the saturation condition,  
\begin{equation}\label{NBH}
 \alpha_{\rm gr}(Q_S) N =1\,, 
 \end{equation}
where $\alpha_{\rm gr}(Q_S)$ is the graviton-graviton coupling 
evaluated at the scale $Q_S$,  
\begin{equation} \label{alphaGR}
 \alpha_{\rm gr}(Q_S) \equiv \frac{Q_S^2}{M_P^2} = (L_PQ_S)^2\,. 
 \end{equation}
Here (in speed of light units) 
$M_P=\sqrt{\hbar/G}$ and $L_P=\sqrt{\hbar \,G}$ denote the Planck mass and length
respectively, with $G$ denoting Newton's constant. 

The physical meaning of Eq.~(\ref{NBH}) can be understood 
from several perspectives. It defines the condition for the $N$-graviton state to be self-sustaining; when this relation is saturated, the kinetic energy of each graviton 
is balanced by the attractive potential energy from the rest of the gravitons -- the self-bound gravitons form the long-lived Black Hole bound-state.  Since the 
mass of this bound-state is $M_{\rm BH} = N\,Q_S$, it is clear from
Eqs.~(\ref{NBH}) and (\ref{alphaGR}) that the wavelength of the constituent gravitons is equal\footnote{\label{f:three}There are factors of $2 \pi$ and $O(1)$ constants that can be accounted for 
in computations in a specific framework. Since the focus here is on the universal features of the CGC-BHNP correspondence, independent of the details of the particular frameworks, we will not attempt to carefully account for these constant factors.} to the Schwarzschild radius of a classical Black Hole 
of mass $M_{\rm BH}$,
\begin{equation}\label{QR}
Q_S^{-1} = R_S = \frac{M_{\rm BH}}{M_P^2} \,.
\end{equation}
This graviton bound state decays slowly through the loss of its constituent gravitons as a result of their rescattering. As we will soon discuss, their  rate of  emission matches 
Hawking's thermal evaporation rate in the semi-classical limit $N\rightarrow \infty,~R_S= Q_S^{-1} = {\rm finite}$.

Another physical interpretation of Eq.~(\ref{NBH}) is that it represents the value of the coupling for which the $N$-graviton state unitarizes the scattering 
amplitude for the center-of-mass energy  $\sqrt{s} =M_{\rm BH}$.   
Thus the formation of a Black Hole in the high energy scattering of gravitons can be understood as the creation of a long lived state of highly occupied ($N\gg 1$) albeit weakly coupled ($\alpha_{\rm gr} \sim 1/N$) gravitons. In this quantum portrait of a classical Black Hole, unitarization occurs at IR wavelengths  $R_S = \sqrt{N} L_P$,  much larger than the UV Planck scale $L_P$. Thus the Schwarzschild radius $R_S$ can be understood as the emergent scale characterizing dynamical unitarization at a fixed point given by Eq.~(\ref{NBH}), for any $N\gg 1$. 
 
 \subsection{Color Glass Condensate in QCD} 
 
 Likewise, for  high energy scattering in QCD, the perturbative unitarization line 
 \begin{equation} \label{NCGC}
 \alpha_S(Q_S) N \, =\, 1\,,
 \end{equation}
 corresponds to an emergent saturation scale $Q_S$. A physical interpretation of this relation is that with the emission of $N$ soft gluons, the phase space density in the Regge limit of QCD becomes large due to bremsstrahlung. The further emission of soft gluons is compensated by the recombination and screening of soft gluons, leading to a saturation of their growth when the occupancy is given by Eq.~(\ref{NCGC}). This gluon saturation phenomenon~\cite{Gribov:1984tu,Mueller:1985wy} is equivalently expressed as 
 \begin{equation}
 \label{eq:gluon-saturation}
 N\equiv \frac{xG_A(x,Q_S^2)}{2(N_c^2-1)\pi R_A^2 Q_S^2} = \frac{1}{\alpha_S(Q_S)} \,,
 \end{equation}
 where $R_A$ is the nuclear radius, $xG_A(x,Q_S^2)$ is the corresponding gluon parton distribution, and $Q_S^2\equiv Q^2$, is the resolution scale of the DIS probe that saturates Eq.~(\ref{NCGC}). In the language of the operator product expansion (OPE), this saturation condition is satisfied when ``all-twist" contributions to the growth of parton distributions are of equal magnitude to the leading twist contribution.
 
The equivalence between perturbative unitarization and close packing is manifest in the CGC EFT, where the dynamics of small gluons is represented by classical gluon field 
configurations with maximal occupancy $A_\mu = O(1/\sqrt{\alpha_S})$ corresponding to solutions of the QCD Yang-Mills equations~\cite{McLerran:1993ni,McLerran:1993ka,JalilianMarian:1996xn}. In this framework, Eq.~(\ref{NCGC})  emerges from a computation of the S-matrix for inclusive DIS scattering off a large nucleus at high energies. This S-matrix satisfies a non-linear renormalization group equation\footnote{This JIMWLK equation describes a hierarchy of n-point Wilson line correlators~\cite{JalilianMarian:1997gr,JalilianMarian:1997dw,Iancu:2000hn,Ferreiro:2001qy} evolving with rapidity. The BK equation is the closed form  equation for the lowest two point correlator in this hierarchy,  obtained in the limit of large $N_c$ and large nuclear size $\alpha_S^2 A^{1/3} \gg 1$.}, the Balitsky-Kovchegov (BK) equation~\cite{Balitsky:1995ub,Kovchegov:1999yj}, describing its  evolution in rapidity Y ($=\ln(1/x)$) with the emission of small $x$ gluons in the dense gluon environment.  
For a given impact parameter, the infrared fixed point of the BK equation, corresponding to the unitarization of the cross-section, defines the emergent scale $Q_S\gg \Lambda_{\rm QCD}$ satisfying Eq.~(\ref{NCGC}), where
  $\Lambda_{\rm QCD}$ is the strong coupling scale. From asymptotic freedom, this of course implies $ \alpha_S(Q_S)\ll 1$ in the Regge limit, where $Q_S(x)$ is the largest scale associated with the $2\rightarrow N$ process of interest.  An important point to note is that ``perturbative unitarization" of the $2\rightarrow N$ cross-section, for fixed impact parameter, occurs at much shorter distances $\sim 1/Q_S\ll 1/\Lambda_{\rm QCD}$ than those corresponding to the nonperturbative confining dynamics of QCD. 
  
\subsection{A digression on the UV/IR nature of gravity} 
 
Before we outline the concrete dictionary mapping the CGC to the BHNP,  we need to clarify the properties 
 that are intrinsic to  the scale dependence of the gravitational interaction. 
It is often assumed that the high energy theory probes short distances. 
This is certainly the case in a class of quantum field theories (QFTs)  in which ``Wilsonian"
UV-completion at any scale can be achieved by integrating
in weakly interacting  degrees of freedom corresponding to shorter and shorter 
wavelengths, with QCD being a paradigmatic example.  

 In such Wilsonian QFTs, there exists  a clear connection between the energy scale and the distance probed by it. Namely, if we wish to probe a distance $L$, we must 
   perform a scattering experiment with a center of mass  energy 
 $E \sim 1/L$.  Therefore in a theory with Wilsonian UV behavior, if the particles that scatter have sufficiently high energies, we can probe arbitrarily short distance scales.

 However, evidently there exists a class of theories \cite{Dvali:2010jz} that differ dramatically from Wilsonian ones, of which gravity is a prominent representative \cite{Dvali:2010bf}. In such theories,  
 above certain 
 scale, the scattering is dominated by formation of  classical 
 objects, consisting of high occupation number of ``soft" quanta. 
 This dynamics effectively prevents the theory from entering into a short distance regime.   This concept was referred to \cite{Dvali:2010jz} as the ``UV-completion by classicalization". 
 
  In gravity, increasing the center of mass energy is no guarantee for probing short distances because of the existence of Black Holes. 
 Instead, the formation of Black Holes suggests that scales shorter than the Planck Length cannot be probed \cite{Dvali:2010bf}. 
 This can be understood already at the semi-classical level and, more 
 profoundly, at the level of multi-graviton scattering amplitudes. 
 
 The semi-classical argument is simple \cite{tHooft:1993dmi}.  As soon as the initial center 
 of mass energy gets localized within its Schwarzschild radius, a Black Hole forms.
 In this way, effectively, the scattering experiment with a center 
 of mass energy $\sqrt{s}$, probes distances $L \sim \sqrt{s}/M_P^2$ rather than 
 $L \sim 1/\sqrt{s}$.  That is, deep UV gravity, via Black Holes, effectively 
  is converted into a deep IR gravity \cite{Dvali:2010bf}.   
 
The Black Hole $N$-portrait  \cite{Dvali:2011aa} offers a microscopic understanding of 
 this phenomenon. In field theoretic language, the process of black hole creation 
  means that the initial center of mass energy of colliding hard quanta gets distributed among 
 a large number $N$ of soft gravitons, with wavelengths  $Q_S^{-1} \sim \sqrt{s}/M_P^2$.
 In this way, the momentum transfer in each elementary vertex 
 is minuscule.  Since the graviton coupling (see Eq.~(\ref{alphaGR})) probed in such an experiment is 
 weak, it is easy to see that it satisfies the saturation
 relation in Eq.~(\ref{NBH}).  Because weak coupling is accompanied by saturation, the physics of this regime is highly nontrivial; in particular, 
 the occupation number of quantum and the entropy of the $N$-graviton state have to be just right for this collective effect to be realized. 
 
Note that the above applies equally to effective field theories of quantum gravity as well as its  embedding in string theory.  
String theory fully accommodates the non-Wilsonian asymptotic behavior of gravity. As long as the  Schwarzschild radius corresponding to the center 
of mass energy is much larger than the string length, Black Hole formation in string theory will proceed exactly as in effective field theories of gravity. 
This expectation is confirmed by computations of multi-graviton amplitudes - supposedly describing Black Hole formation
- which agree when computed in both QFT and in stringy regimes \cite{Dvali:2014ila}.  
 
 Our above discussion suggests that while one can say that 
a Black Hole state probes gravity at very high energies, it would be a mistake to call it a UV-regime of gravity.  The relevant scale (which is an IR scale) is set by the frequencies of the constituents and not by the overall energy of the system. 

 It is also important to not confuse the role of the Planck scale, which represents the strength of graviton coupling. 
In other words,  as we will discuss further, it has to be viewed as the graviton decay constant. 
 As pointed out in \cite{Dvali:2020wqi}, when performing the mapping  between a black hole and a generic saturated system, 
 the Planck mass must be mapped on the decay constant $f$ of the Goldstone boson of the spontaneously broken 
 Poincar{\'{e}} symmetry in that system.  This Goldstone boson must be universal since any saturated system spontaneously breaks Poincar{\'{e}} symmetry. 
  
 \subsection{Dictionary mapping gravity in the UV to QCD in the IR} 
 
 It is now straightforward to establish the dictionary between the BHNP and the CGC. Both states are uniquely determined 
   by the corresponding saturation equations in Eq.~(\ref{NBH}) and 
   Eq.~(\ref{NCGC}) respectively. In both cases, the saturation scale $Q_S$ determines the typical 
   momenta of constituent gravitons and gluons respectively. With this dictionary, a region of space 
   of radius $R_S = Q_S^{-1}$ of the CGC can be mapped on to a Black Hole 
   of  radius determined by the saturation of the graviton 
   condensate.    
    
 Key features of the dynamics mapped between the two systems exhibit  UV/IR relations that become apparent 
when we consider the behavior of the underlying theory in these two regimes.
 In gravity, the UV scale is given by the Planck scale $M_P$ and in QCD the corresponding IR scale is $\Lambda_{\rm QCD}$.

The significance of these scales is of course that the respective theories enter into the strong coupling regimes when the physics is sensitive to them.
On the other hand, since the unitarization relations in Eqs.~(\ref{NBH}) and 
 (\ref{NCGC}) are saturated at weak coupling,  the saturation scale $Q_S$ is very far at large $N$ from the corresponding strong coupling scale:
 \begin{eqnarray} \label{HScales} 
{\rm for~BHNP}: ~~ Q_S^{-1} & \gg &  M_P^{-1} \,, \\ \nonumber 
{\rm for ~CGC}: ~~ Q_S & \gg &  \Lambda_{\rm QCD} \,.
\end{eqnarray}
in the UV and IR respectively. As we will soon discuss,  the saturation scale in both cases 
 exhibits the same relation with the decay constant $f$ of the Goldstone
  boson of the broken Poincare symmetry resulting from the creation of these high occupancy states:   
   \begin{eqnarray} \label{HGscales} 
{\rm for~BHNP}: ~~ Q_S^{-1} & = & \sqrt{N} f^{-1} \,, \\ \nonumber 
{\rm for ~CGC}: ~~ Q_S^{-1}  & = & \sqrt{N} f^{-1}  \,.
\end{eqnarray}
 This scale $f^{-1}$ is equal to $\sqrt{\alpha}$ times the screening length of the corresponding degree of freedom (graviton or gluon). For a black hole, this corresponds to  $f=M_P$. 

 The lesson here is that the form of the correspondence between BHNP and CGC depends 
on the resolution scale $Q$ at which we  probe these states. When 
$Q=Q_S$, the correspondence is direct, and driven by the dynamics of saturation at weak coupling in both theories, with $\alpha = 1/N \ll 1$. 

However UV/IR type differences
 emerge whenever we discuss the relations between the saturation momentum $Q_S$ and 
 the strong coupling scales in the two theories,  once we move to much shorter distances from $Q_S^{-1}$ in gravity or much longer distances in QCD.
   Hence in order to maintain their qualitative similarities while changing the scale, 
 we must synchronize the signs in the variations of the couplings in the two theories. That is, a motion towards the UV in the BHNP must be mapped 
 on the motion towards the IR in the CGC and vice-versa. Of course, away from the saturation point, the correspondence is 
  lost in general. 
  However it is conceivable that, on the weak coupling side, features of this correspondence may benefit from Lipatov's EFT for reggeized gravitons and gluons~\cite{Lipatov:1982vv,Lipatov:1991nf}, the previously noted BCJ relations~\cite{Bern:2008qj} and their extension to  classical ``double copy" relations~\cite{Monteiro:2014cda,Goldberger:2016iau}. 
 
 \section{Connecting Entropies in the BHNP and in the CGC} 

The connection between the micro-state entropies of CGC and BHNP 
can be summarized by the fact that both satisfy Eqs.~(\ref{USB}) and (\ref{Gf}).  
 According to 
 \cite{Dvali:2020wqi}, (see also, \cite{Dvali:2019jjw, Dvali:2019ulr})
  the above equalities
 are universally satisfied by critically packed states (defined by Eq.~(\ref{NA})) that saturate the unitary bound on entropy, resulting into the combined saturation relations:  
 \begin{equation} \label{universal} 
S  = N = \frac{1}{\alpha(Q)} = {\rm Area}\, \times f^{2} \,,  
\end{equation} 
where, as before, $f$ is the decay constant of the Goldstone boson 
of broken Poincar{\'{e}} symmetry.  
  We will first reproduce the general argument of \cite{Dvali:2020wqi} for the validity of this relation and then specialize the discussion to Black Holes and the CGC. 
      
\subsection{General Argument} 

Consider a critically packed state of occupation  number  $N=\alpha^{-1}$ of quanta with momentum $Q_S$
and volume $R_S^3 = Q_S^{-3}$.  We shall assume that the entropy 
is given by Eq.~(\ref{USB}). 
 Let us now show that the same entropy is also equal to the area of the region (= $4 \pi R_S^2$), in units of a Goldstone decay constant $f$. The Goldstone  modes in question originate from the fact that any critically packed condensate spontaneously breaks symmetries both in coordinate and in relevant internal spaces.   As discussed in Appendix A, the order parameter of this breaking is given by $N\,Q_S^2$ and the decay constant of the canonically 
 normalized Goldstone mode is
 \begin{equation} \label{f}
 f = \sqrt{N}Q_S \,.
 \end{equation}
 From this relation\footnote{This behaviour was explicitly demonstrated in \cite{Dvali:2019ulr, Dvali:2019ulr, Dvali:2020wqi} for several field theory examples of extended objects such as solitons, instantons and baryons in large-$N_c$ QCD. }, it is clear that the entropy  in Eq.~(\ref{universal}) of the region filled with the condensate  is nothing but 
 the area of that region (Area $= 4\pi\,R_S^2 = 4 \pi \,Q_S^{-2}$) measured in units of $f^2$. 
 
 Notice, as a byproduct, the above relation implies the saturation 
 of the Bekenstein bound on entropy.  
Indeed, using the saturation relation in Eq.~(\ref{USB}) and the 
critical packing relation in Eq.~(\ref{NA}) we can write, 
\begin{equation} \label{Sregion} 
  S= \alpha^{-1} =  N\,Q_S\,R_S = E\,R \, ,
 \end{equation} 
 This last expression is nothing but the energy of the condensate, $E = N Q_S$ times its size $R_S$. We have therefore  
 the well-known expression~\cite{Bekenstein:1980jp} for the Bekenstein entropy\footnote{As per our remarks in footnote \ref{f:three}, we omit the factor of $2\pi$ in this expression.}.  
 
\subsection{Entropy of a Black Hole} 

We shall now apply the above reasoning to the BHNP. The observation that the Bekentein-Hawking entropy of a black hole
 is equal  
to the inverse graviton coupling evaluated at the momentum transfer $Q_S = 1/R_S$ (or equivalently, 
to the number $N$ of Black Hole constituents) was made in \cite{Dvali:2011aa}. 
 Indeed, it is obvious from Eq.~(\ref{alphaGR}) that the coupling  ${\alpha_{\rm gr}}$ is nothing but the inverse of the Black Hole area in Planck units.
Thus the Black Hole entropy is not just an area but also the inverse 
 of the gravitational coupling:  
\begin{equation} \label{BHentropy} 
S_{\rm BH}  = \frac{1}{\alpha_{\rm gr}} =\frac{4\pi R_S^2}{4 L_P^2} \,.
\end{equation} 
This is a particular form of the more general relation in Eq.~(\ref{universal}). 
This way of understanding the Black Hole entropy connects {\it the geometrical notion of an area to the  
quantum field theoretic concept of the graviton coupling}.  

Further, the BHNP reveals the microscopic meaning of the 
simultaneous saturation of the entropy bound and unitarity. With regard to the latter,  as noted previously, the S-matrix for Black Hole
formation in the scattering of two trans-Planckian quanta with center of mass energy $\sqrt{s} \gg M_P$ is that of $2\rightarrow N$ scattering, where 
the final $N$ graviton quanta have momenta given by 
$Q_S = \frac{\sqrt{s}}{N} = R_S^{-1}$.  A detailed computation of this process, both in quantum field theory as well as in string theory, was 
performed in \cite{Dvali:2014ila}. At large $N$, both computations 
at the  Black Hole formation threshold give the following rate for the $2\rightarrow N$ process: 
\begin{equation} \label{2toN} 
 \Gamma_{2\rightarrow \rm{BH}} \sim 
  {\rm e}^{-\frac{1}{\alpha_{\rm gr}}+ S_{\rm BH}} 
 \end{equation} 
where $\alpha_{\rm gr} = \frac{Q_S^2}{M_P^2} = (R_SM_P)^{-2}$. The 
exponential factor ${\rm e}^{S_{\rm BH}}$
 comes from the summation over Black Hole micro-states. 
 The  form of the rate in Eq.~(\ref{2toN})  highlights the role of  Eq.~(\ref{BHentropy})
in the saturation of unitarity by Black Holes.   
The bottomline is that the Black Hole entropy is equal to the inverse 
graviton-graviton coupling at the point where unitarity is saturated, in full accord with Eq.~(\ref{universal}). 

Finally, as discussed in \cite{Dvali:2019ulr, Dvali:2020wqi},
 the Black Hole entropy given by Eq.~(\ref{BHentropy}), 
 is a particular example of the much more general expression
 in Eq.~(\ref{Gf}) for the micro-state entropy of a saturated system.   
As already noted, this expression states that the maximal 
entropy of an object is equal to the area in units of the  
Goldstone decay constant.

 In the case of a Black Hole, as discussed in
 \cite{Dvali:2020wqi} and in Appendix A, the Goldstone mode is 
the graviton itself. Indeed, the saturated state of the graviton condensate 
  breaks Poincar{\'{e}} invariance spontaneously. The order parameter of this breaking 
  is $N Q_S^2 = N/R_S^2$ and the decay constant of the corresponding  
 Goldstone mode in Eq. (\ref{f}) is equal to 
$f = \sqrt{N}/R_S =M_P$, which is nothing but the Planck mass.  
This result makes  physical sense because on the one hand, the Planck mass is the graviton decay constant, and on the other, the  Black Hole is a graviton state of maximal occupation number 
so that the expectation value of the graviton field 
is $M_P$.  The last statement is clear already from 
the classical theory, as this is the value attained by the 
Schwarzschild  metric tensor
relative to the Minkowski metric. 

 \subsection{Entropy of the CGC} 
  
We now wish to argue that Eq.~(\ref{universal})  is also satisfied by the CGC. We shall apply the generic reasoning 
of \cite{Dvali:2020wqi} 
which shows that states saturating unitarity must also saturate the entropy bounds Eqs.~(\ref{USB}) and ~(\ref{Gf}),   
and therefore Eq.~(\ref{universal}). 

As we noted previously, the CGC is a state of maximal occupancy in the proton that is excited in DIS  or hadron-hadron scattering in the high energy Regge limit of QCD. 
In this $2\rightarrow N$ scattering, a high energy  probe effectively interacts with a lumpy deconfined configuration of gauge fields ($A_\mu\sim O(1/\sqrt{\alpha_S})$) with a typical correlation length $R_S\sim 1/Q_S \ll 1/\Lambda_{\rm QCD}$~\cite{Gelis:2010nm}. From the ``perspective" of the probe, the wee gluons in the boosted proton are Lorentz contracted 
to a distance\footnote{Note that since $Q_S\equiv Q_S(x)$, this picture is invariant under boosts. } $1/Q_S$, in contrast to the high energy valence parton modes that are contracted by $2\,R_p/\gamma$, with $\gamma \sim \sqrt{s}/m_p$ the Lorentz factor of the valence parton modes,  and $R_p$,  $m_p$  the proton mass and radius respectively.

 With this in mind, let us consider the CGC spatial volume
 $V_{\rm CGC} = R_S^3 = Q_S^{-3}$ of saturated wee gluon modes.  The energy contained in this region, as seen by the high energy probe, is then simply
 \begin{equation} \label{ECGC}
 E_{\rm CGC} = N\, Q_S \,. 
 \end{equation}  
 Now just as in the Black Hole case,  we can think of the CGC as an approximately classical macro-state representing a large multiplicity of  quantum ``parton" micro-states. 
 In other words, these micro-states  are indistinguishable classically but represent  distinct quantum states, which constitute the entropy $S_{\rm CGC}$ of the CGC in a region of rapidity 
 $Y= \ln(1/x)$ relative to the probe.
  
 As we noted previously,  explicit weak coupling computations show that the S-matrix for $2\rightarrow N$ scattering is saturated for fixed impact parameter 
 in the Regge limit of QCD.  
 The physical meaning of this key piece of information is as follows.
The statement that the CGC  saturates unitarity (for a given impact parameter) means that its micro-states account for the {\it entire portion}  of the Hilbert space 
describing the complete  set of QCD states localized within the region $V_{\rm CGC}$  with energy $E_{\rm CGC}$.  Indeed, if QCD matter could exist  
 in other possible states within the same region $V_{\rm CGC}$, this would imply that our statement about the CGC saturating unitarity would be false since such additional non-CGC states would be formed in the same scattering process with non-zero probability, thereby violating the unitarity of the CGC state.  
 
 The scaling of the CGC entropy as in Eq.~(\ref{USB}) can be understood as follows.  When $N$ is very large, the CGC can be treated as 
 a classical condensate. Thus when forming CGC, say, in the scattering of gluons in the Regge limit, we are effectively 
 dealing with a transition process in which a classical object is formed 
 from an initial quantum state. In other words, we are dealing with a typical classicalization process in which a formation probability  
 of a classical condensate matches the one of a  $2\rightarrow  N$ quantum scattering \cite{McLerran:1993ni, Dvali:2011th}. Correspondingly, the probability 
 can be understood in two different ways which, for consistency, must give the same result.  
  
 The first way is to think about the process semi-classically. That is, we assume that the bulk of the suppression in forming the state 
 comes from a classically forbidden Euclidean trajectory. This aspect is similar to any text-book  semi-classical computation of tunneling, a  classic example being Coleman's 
 treatment of false vacuum decay \cite{Coleman:1980aw}.  
 
  However in the present case there is a slight technical complication. Ordinarily, in a false vacuum decay, both initial and final 
  configurations are classical. Correspondingly, the classical Euclidean 
  trajectory, a so-called ``bounce", is relatively straightforward to
  find by solving the classical equations of motion in Euclidean space.   
   Such a luxury does not exist in the present case, since the initial 
   two-gluon state cannot be described classically. 
   However this is not a serious obstacle, since  the physics of the suppression remains unaltered and can be reliably estimated
even without having a full-fledged saddle point classical solution
in Euclidean time. 

That is, even in the absence of such a solution, we can still think of the process as ``happening" in the Euclidean time and thus being described by some intermediate Euclidean trajectory.  This virtual trajectory, close to the initial two-gluon state, cannot be described classically. 
 However, it is reasonable to assume that this initial quantum portion contributes very little in the Euclidean action, as compared to the part of the trajectory 
   that is well described classically.  The minimal action of this classical 
  portion can be estimated without having an exact solution at hand. 
  
 Since $E_{\rm CGC} = N\,Q_S$ 
 is the energy of the CGC, and $R_S$ is the typical ``duration" of the  Euclidean time required for its formation, a reasonable guess
 for the minimal action of the relevant Euclidean trajectory is  $N\,Q_SR_S = N$. 
 
 The alternative way \cite{Dvali:2011th}
 to arriving at the same expression is by thinking of the Euclidean action needed to create $N$ particles of momentum 
$Q_S$ from an initial two-particle state. The action per particle is the product of the de Broglie  
  wavelength and energy, $Q_S\,R_S \sim 1$. Therefore, for  $N$ particles this gives the total minimal action $\sim N$. 
  
   The bottom line in the above estimates is that the probability of creating each micro-state of CGC is suppressed by
   \begin{equation}  
   \label{EXPN} 
           {\rm e}^{-N} = {\rm e}^{-\frac{1}{\alpha_S}}\,.  
 \end{equation}  
 This expression is equivalent to the universal suppression factor for a unitary $2 \rightarrow N$ transition to a generic $N$-particle micro-state, derived from general consistency considerations\footnote{That derivation \cite{Dvali:2020wqi}, which will not be reproduced here, does not rely on a particular assumption  about the structure of the theory.  The matching with it therefore provides a consistency check of the above estimate.}.
 
 Explicit computations~\cite{Ayala:1995hx,JalilianMarian:1997dw,Iancu:2000hn,Gelis:2008rw} of small fluctuations on top of the classical CGC solution demonstrate that the $\alpha_S$ suppression is compensated by the phase space in rapidity for gluon emission; such contributions are $\alpha_S Y\sim O(1)$ for $Y\sim 1/\alpha_S$. The resummation of such contributions, in the classical background, to all orders, leads to the BK/JIMWLK equations, which as noted previously, unitarizes the $2\rightarrow N$ scattering amplitude.
 
 The fact that the CGC saturates unitarity then implies that the number of micro-states ($\sim e^{S_{\rm CGC}}$) must compensate the suppression in Eq.~(\ref{EXPN}). Hence it follows that the entropy of the CGC in the volume $V_{\rm CGC}$ must be
\begin{equation} \label{SCGC}
S_{\rm CGC} = \frac{1}{\alpha_S}\, \equiv N  \,.
\end{equation}
This demonstrates that the entropy of the CGC satisfies the first equality in the universal relation in Eq.~(\ref{universal}), or equivalently, 
 saturates the bound in Eq.~(\ref{USB}). 
 
 Of course, by the generic reasoning already given above, as an added bonus,  Eq.~(\ref{SCGC}) automatically implies the saturation of the Bekenstein bound, 
  \begin{equation} \label{BCGC}
S_{\rm CGC} = E_{\rm CGC} \,R_S \,. 
\end{equation}

We will now argue that (just as for the Black Hole case) the entropy does not scale with the volume of the CGC  but instead as its area 
in units of a decay constant $f$ of the Goldstone boson corresponding to the breaking of Poincar{\'{e}} invariance by the CGC. This follows from the generic argument of \cite{Dvali:2020wqi} 
leading to Eq.~(\ref{universal}), also discussed in Appendix A. Here we shall apply this reasoning to the specifics of the CGC.   

In QCD, partons are characterized by hard transverse momenta $p_\perp\sim Q$, where $Q$ is the (UV) momentum resolution scale of the probe. The longitudinal momenta on the other hand (depending on the $x$ values probed) can range from $\Lambda_{\rm QCD}$ to $Q$ in QCD's Bjorken limit, where the parton concept is robust. 

In the Regge limit, the picture changes qualitatively due to the emergent saturation scale $Q_S$. For momentum modes below this scale, the parton concept is not viable and the corresponding micro-states are indistinguishable from a classical field. A simple computation in the CGC shows that the typical transverse momentum is of order $Q_S$. Interestingly, the softest longitudinal momentum modes that carry any field strength also typically have momenta $O(Q_S)$. In other words, as the proton is boosted, the valence modes seen by a probe shrink by the Lorentz factor; however, the wee modes shrink as well, albeit more slowly and with a rate computable in the CGC, and their field strengths in the longitudinal direction are localized on the lightcone at $x^- \sim 1/Q_S$, or $p^+\sim Q_S$. 

There is however an intermediate hard scale between $Q_S$ and $Q$ which characterizes the transition from a gapped dilute gas ``parton gas"  picture to that of the overoccupied CGC classical field. Partons below this scale are screened or, in QCD language, are ``shadowed"; in other words, many-body (or higher twist, in OPE language) effects become important. Just as for a thermal gas, the screening of parton modes breaks the translational invariance of the parton gas; the Goldstone scale $f$ corresponding to this breaking of Poincar{\'{e}} invariance
is exactly analogous to that in the BHNP. Indeed, as argued in 
\cite{Dvali:2020wqi}, on general grounds, this scale is of order $f\sim \sqrt{N} Q_S$ for any classical ``soliton-like" lump with
characteristic momentum $Q_S$ and the occupation 
number $N$.  We shall discuss these arguments further in Appendix A. 
Thus, 
\begin{equation}
S_{\rm CGC} = \frac{1}{\alpha_S} = N = f^2\, {\rm Area}\,,
\end{equation}
since ${\rm Area}\propto R_S^2$. 
 We thus arrive at a universal relation (\ref{universal}) 
 characteristic of a generic saturates system \cite{Dvali:2020wqi}. 

To summarize, dynamical screening is how the CGC restores unitarity; this breaks Poincar{\'{e}} invariance, leading to area scaling of the entropy as opposed to volume scaling, with the scale set by the Goldstone constant $f$. 
As is evident, the above expressions are identical to the ones satisfied by the Black Hole entropy and possess the same microscopic meaning as revealed by the quantum $N$-portrait\footnote{This unitarity limit of maximal screening in the high energy QCD literature is very appropriately referred to as the ``black disc" limit-for a nice discussion of the different regimes, see~\cite{Kancheli:2020xho}.}. 

For QCD dynamics in the Regge limit, from solutions of the ``mean field" CGC BK equation\footnote{The parton distributions corresponding to the solutions of this equation indeed have the structure of a propagating soliton-like wavefront~\cite{Munier:2003vc,Munier:2003sj}. The kinematic region between $Q_S$ and $p_{\perp, {\rm shdw}}$ is often referred to as the ``geometrical scaling" or ``leading twist shadowing" regime in the small $x$ QCD literature, and there are strong arguments that small $x$ DIS data from HERA can be interpreted in this light~\cite{Stasto:2000er,Iancu:2002tr}.},  one can deduce this scale to be closely related to $p_{\perp, {\rm shdw}}^2 \sim Q_S^4/\Lambda_{\rm QCD}^2$~\cite{Iancu:2002tr}. 
The latter can be expressed as $p_{\perp, {\rm shdw}}^2 \sim f^2 \,\alpha_S\, n_D$, where $n_D= Q_S^2/\Lambda_{\rm QCD}^2$ is the number of domains in the proton. 

Since the entropy of the CGC is given by its area and not its volume, the maximal total entropy  of wee partons released from scattering off the proton is 
given by the ratio of the transverse area of the proton $\propto \frac{1}{\Lambda_{\rm QCD}^2}$ to that of the CGC, 
\begin{equation} \label{STOTAL}
S_{\rm tot} =
  \frac{1}{\alpha_S} \frac{1}{\left(R_S \, \Lambda_{\rm QCD}\right)^2} = n_D \,S_{\rm CGC} 
 \,,
\end{equation}
where we employed 
Eq.~(\ref{SCGC}). 
We assumed here that, since $R_S$ dynamically screens color charge at larger distances, the  micro-states of each of the $n_D$ domains, given by the ratio of the area of the proton ($\sim 1/\Lambda_{\rm QCD}^2$)  to that of a domain ($R_S^2$), 
are not correlated.  Hence  the total entropy of the proton is given by the number of domains times the entropy of each. 

Note that the occupation number $N$ (implicitly through the running of the coupling) as well as $n_D$  and $Q_S$ (explicitly) depend on the intrinsic transverse size of the proton $\Lambda_{\rm QCD}$, so the relation between $N$ and $n_D$ is an additional input that must self-consistently satisfy the relations in 
in Eq.~(\ref{NCGC}), Eq.~(\ref{SCGC})  and (\ref{STOTAL}). In other words, the universality we posit is between the Black Hole state and a lump of CGC and therefore in principle 
 independent of the number of saturated lumps $n_D$ that can fit within the proton. Of course, the running of the coupling in QCD ensures that the two are not entirely independent of each other. 

When entropy is released in the lab frame of the scattering, it is appropriate to describe it in terms of the rapidity variable, which adds additively under boosts between frames\footnote{Indeed, as argued in \cite{Berges:2017zws,Berges:2017hne}, it is only meaningful to rigorously define this entanglement entropy for a finite interval of rapidity.}. In the bremsstrahlung process, on average, one parton is released in $1/\alpha_S$ units of rapidity: $\alpha_S \,dY\sim 1$. Hence the boost invariant distribution of entropy 
in the scattering process is given by\footnote{It would be nice to compare our estimates of the micro-state entropy 
of CGC  (which is determined by the number of nearly-degenerate 
pure micro-states)  with the entanglement entropy in the CGC using the techniques discussed in \cite{Berges:2017hne,Kovner:2015hga,Duan:2020jkz,Armesto:2019mna}. Since the typical scaling of the entanglement  entropy in local field theories is given by the area 
in units of the cutoff of the theory, we expect that the two must match, at least when entanglement entropy is computed in the EFT of the Goldstone modes.  Indeed, the natural cutoff of this effective theory is the Goldstone decay constant $f = \sqrt{N} Q_S$.  Thus, the entanglement entropy of the CGC in a volume $Q_S^{-3}$, must scale as area in units of $f$, thereby matching the general bound in Eq.~(\ref{Gf}).}
{\bf 
\begin{equation}
\frac{dS_{\rm tot}}{dY} = n_D (\alpha_S) \,.
\end{equation} 
} 

There have been several discussions in the literature on the entropy of the CGC and the proton at small $x$~\cite{Kutak:2011rb,Muller:2011ra,Peschanski:2012cw,McLerran:2014hza,Kovner:2015hga,Kharzeev:2017qzs,Hagiwara:2017uaz,Duan:2020jkz,Armesto:2019mna}. 
Comparisons of our results to these are not straightforward since several employ different definitions of the entropy or address different kinematic regions in high energy 
scattering\footnote{We would like to thank V. Skokov and Z. Tu for discussions pertaining to these issues.}. The closest treatment to ours is that of 
\cite{Kharzeev:2017qzs} which likewise addresses the Von Neumann entropy obtained from the counting of the quantum micro-states of the CGC. In deeply inelastic scattering (DIS), the relevant quantity is the entropy of a lump of CGC ($n_D=1$) given in Eq.~(\ref{SCGC}) that unitarizes DIS cross-section in Regge asymptotics:
\begin{equation}
S^{\rm DIS} = \frac{1}{\alpha_S(Q_S)} \,\, \rightarrow \,\, \frac{dS^{\rm DIS}}{dY} = {\rm constant} \, .
\end{equation}
This result is consistent with that of \cite{Kharzeev:2017qzs}. In contrast, in hadron-hadron (h-h)collisions, since several lumps of CGC are produced, the entropy released is given by 
\begin{equation}
S^{\rm h-h} = n_D\,\frac{1}{\alpha_S(Q_S)} \,\,\rightarrow \,\,  \frac{dS^{\rm h-h}}{dY} = n_D\,.
\end{equation}
This estimate is consistent with previous estimates of the Glasma entropy~\cite{Mueller:1999fp,Krasnitz:2000gz,Lappi:2006fp} at $t = 1/Q_S$ if there is a solution to the self-consistency relation\footnote{Note that this assumption would also give us $p_{\perp,{\rm shdw}}^2 = f^2 \alpha_S n_D \sim f^2$.} $n_D\sim N=1/\alpha_S$. 

For collisions of heavy-nuclei, for rapidities $Y \leq \ln^2(A^{1/3})$, this estimate should be multiplied by the atomic number $A$ since $Q_{S,A}^2 = \Lambda_{\rm QCD}^2 A^{1/3}$~\cite{Gribov:1984tu,Mueller:1985wy,McLerran:1993ni} and $R_A \propto A^{1/3}$. However at asymptotically high energies corresponding to $Y\gg  \ln^2(A^{1/3})$, $Q_S^2$ becomes independent of $A$~\cite{Mueller:2003bz}; in this asymptotics, the entropy will scale as the area of the nucleus $R_A^2\propto A^{2/3}$. 

\section{Goldstone Origin of Entropy and the Time-Scale of 
Information-Retrieval} 

In this section, following \cite{Dvali:2020wqi}
\cite{Dvali:2019jjw, Dvali:2019ulr}, we shall give a Goldstone interpretation of the quantum information carrying modes that are responsible for the micro-state entropy of a generic system.   This provides a deeper understanding of the appearance of the Goldstone decay constant $f$ in the area law scaling of the entropy
in Eq.~(\ref{Gf}).  The Goldstone interpretation of entropy was originally introduced in the context of BHNP and the key essence was also illustrated 
  for a critically packed  condensate of bosons in \cite{Dvali:2015ywa}.

As a further step,  it was shown in \cite{Dvali:2020wqi,Dvali:2019jjw, Dvali:2019ulr}
that 
the micro-state entropy of  systems in ordinary gauge theories, 
and its connection with  
unitarity, can be understood and explained in terms of the Goldstone 
phenomenon.

 This Goldstone formulation 
allows one to derive a universal expression \cite{Dvali:2020wqi}, presented below, for the time-scale
required for retrieving quantum information stored 
in micro-states of the system.  
In particular, this knowledge was applied to saturated 
multi-particle states such as lumps of gluons or baryons in 
QCD with a large number of colors. 
 Due to the universality of  the Goldstone phenomenon, the same should be 
 applicable to the CGC.  
  
 Let us consider the critically packed state 
 (Eq.~(\ref{NA})) of some gauge degrees of freedom.   We shall assume that 
 the state is self-sustained, meaning that its ``life-expectancy" 
 is a growing function of $N$. In certain cases, the configuration may 
 be exactly stable due to conserved quantum numbers.  For example, 
this can be the case for topological solitons and baryons (Skyrmions). 
   For such a system, the infinite-$N$ limit is well-defined and, in this limit, 
   corresponds to a classical background configuration 
   which we shall generically refer to as a ``lump".

 The first key point of \cite{Dvali:2019ulr, Dvali:2020wqi} is that 
 such a field configuration necessarily breaks spontaneously 
 a set of global symmetries. As already discussed, among 
 the broken symmetries is Poincar{\'{e}} invariance, as the condensate transforms non-trivially under it.
  However in addition, there exist broken global symmetries 
 that parameterize the embedding of the lump into the group space.  
 
 At infinite $N$, the notion of a spontaneous breaking of symmetry 
 is exact and the lumps with different internal orientations
 can be viewed as different ``vacua". They are connected by the 
 broken global symmetry transformations and are strictly  degenerate in energy.   Correspondingly, there 
 exist associated Goldstone modes. These modes are gapless at 
 $N=\infty$.    At finite $N$, the Goldstones remain gapless, at least 
to leading order in $1/N$. 
 
 The lumps corresponding to different orientations form different {\it micro-states}  of the same {\it macro-state}. The resulting micro-state entropy is defined as the log from the number of independent 
 orientations $n_{\rm st}$,  
 \begin{equation} \label{Sn} 
    S = \ln(n_{\rm st}) \,.  
 \end{equation}   
 The different orientations are obtained from one another by the excitations of the Goldstone modes.  
  This fact makes it transparent that the Goldstone modes are the carriers of the micro-state entropy.

 We next present a universal expression for the time scale 
 required for the retrieval of the quantum information, derived in 
  \cite{Dvali:2020wqi}. 
 Since quantum information is stored within the Goldstone modes, 
  the retrieval of this information requires the read-out of this Goldstone 
 state which in turn necessarily involves interactions with the Goldstone modes.    
 Thus the processing of quantum information cannot happen on time scales shorter than the
 Goldstone interaction time given by 
 \begin{equation} \label{tG}
  t_{\rm Goldstone} = \frac{1}{\Gamma_{\rm Goldstone}}  \sim f^2 R_S^3 = N R_S,  
 \end{equation} 
 where  $\Gamma_{\rm Goldstone}$ is the Goldstone interaction rate (the cross-section times the density of such modes) for 
 the characteristic momentum transfer scale $Q_S = 1/R_S$. This expression therefore sets an universal lower bound on the time scale required for retrieving quantum information 
 stored within the system. 
 The Goldstone argument explains why this time-scale is macroscopic and is determined by the number of constituents $N$. 
 From the saturation relation (Eq.~(\ref{universal})), this time-scale can be expressed as 
   \begin{equation} \label{tAl}
  t_{\rm Goldstone} = \frac{R_S}{\alpha} 
 \end{equation} 
 As  noted in  \cite{Dvali:2020wqi} for the particular case of the BHNP, the expression in Eq.~(\ref{tG}) and Eq.~(\ref{tAl}) 
  reproduce the time-scale proposed by Page~\cite{Page:1976df} for the time required to retrieve information from a Black Hole. 
  These time-scales also agree with the half-decay time of a Black Hole and, as we shall now discuss, with its 
quantum break-time \cite{Dvali:2013vxa}.        
   
\section{Comparing time-scales for Black Hole decay and for Glasma evolution to a Quark-Gluon Plasma}

\subsection{Semi-classical and quantum portraits of Black Hole life time}

In Black Hole physics,  an extremely important time-scale can be expressed as follows, 
\begin{equation}
\label{time}
  t_{\rm BH}  =  R_S \frac{R_S^2}{L_P^2}  = N\,R_S = \omega^{-1}=
   \alpha_{\rm gr}^{-1}R_S \,,
  \end{equation} 
where, as we will now discuss,  the four equivalent ways of writing this time-scale will be convenient for making transparent the physical meaning of each.   
 
The first relation $t_{\rm BH}  =  R_S \frac{R_S^2}{L_P^2}$ is physically meaningful for two reasons, neither of which requires any knowledge of the microscopic theory. 
The first macroscopic interpretation of $t_{\rm BH}$ is that of  the half-decay time of a Black Hole due to its semi-classical evaporation via Hawking
radiation~\cite{Hawking:1974sw}. The latter can easily be obtained from the Stefan-Boltzmann law applied  to Hawking's  thermal evaporation relation, 
\begin{equation} \label{StefanB}
 \frac{dM_{\rm BH}}{dt} = -T_H^2 \,,
\end{equation} 
where  $T_{\rm H} = R_S^{-1} = M_P^2M_{\rm BH}^{-1}$ is the Hawking temperature.  We should note that the microscopic theory of the BHNP
reveals that  Eq.~(\ref{StefanB}) is valid until the half-decay time. Beyond this point, the semi-classical analysis can no longer be trusted. 

The second  macroscopic interpretation of $t_{\rm BH}$, as noted previously, is that of Page \cite{Page:1976df}, who argued that $t_{\rm BH}$ is the minimal time required for an observer to be able to  resolving outgoing information at an order one rate.  

The other three ways of writing $t_{\rm BH}$ in Eq.~(\ref{time}) correspond to three different microscopic interpretations of the same physics that is provided by the BHNP.
Two of these clarify our microscopic understanding of the above interpretations {\it a la}  Hawking and Page respectively and reveal
 underlying structure not captured by the semi-classical theory. The final microscopic relation is noteworthy because it provides an interpretation of $t_{\rm BH}$ 
that has no equivalent counterpart within the semi-classical picture of a Black Hole. 

We shall now consider each  of these three microscopic relations separately. When we write Eq.~(\ref{time}) as  
$t_{\rm BH} = N\,R_S$, it describes the time during which the Black Hole graviton condensate loses an $O(1)$ fraction 
of its initial $N$ constituents. Indeed according to the BHNP, for each Hawking emission time of $\sim R_S$, the condensate 
on average loses one constituent, significantly depleting over $\sim N$ emissions.  Hence this form of $t_{\rm BH}$ 
provides a microscopic explanation to the half-life of a Black Hole obtained in
Hawking's semi-classical theory. 

 The second microscopic relation in  Eq.~(\ref{time}), $t_{\rm BH} =\omega^{-1}$, equates the l.h.s  to the inverse of $\omega \sim \frac{1}{S_{\rm BH} R_S} 
 \sim   \frac{1}{N R_S}$, which represents the characteristic  energy gap of information-carrying Bogoliubov-Goldstone modes
 of the graviton  condensate \cite{Dvali:2012en, Dvali:2015ywa}. 
  This relation then gives a transparent microscopic interpretation of Page's time:  Information encoded in a quantum degree of freedom cannot be 
 resolved faster than the inverse of its excitation  energy. As we discussed already, Eq.~(\ref{time}) is particular example of  Eq.~(\ref{tG}), stating that 
information-retrieval time from an arbitrary system is given the interaction time 
of information carrying Goldstone modes \cite{Dvali:2020wqi}.   
 
The last equality in Eq.~(\ref{time}), $t_{\rm BH} = \alpha_{\rm gr}^{-1}R_S \equiv t_Q$, provides us  
 with a qualitatively different microscopic meaning of  $t_{\rm BH}$ being the {\it quantum break time}.  
 This notion has no analog in semi-classical theory and describes a time scale beyond which the quantum evolution of the Black Hole 
 departs fully from the classical one \cite{Dvali:2013vxa}. As argued in \cite{Dvali:2017eba},  this time scale is characteristic of critically packed systems for which it has the following simple form, 
\begin{equation} \label{Qtime}  
t_Q = \frac{1}{\alpha\, Q_S} \,,
\end{equation} 
where $Q_S = 1/R_S$ is the characteristic momentum scale and 
$\alpha$ is the coupling.  For the 
BHNP,  $\alpha=\alpha_{\rm gr}$ and $R_S$ is the gravitational radius.
 
 Again, the similarity between the quantum break-time Eq.~(\ref{Qtime})  and minimal time of information-retrieval 
 Eq.~(\ref{tAl}) has a deep physical meaning as both processes are controlled by the strength of quantum interactions \cite{Dvali:2020wqi}.   

However the phenomenon is very general 
and takes place even when more than one scale is present. In such cases, the parametric dependence on 
$\alpha$ may change. For example, it was shown in
\cite{Dvali:2013vxa} that in the presence of classical time evolution, or 
Lyapunov instabilities, the quantum break time can be significantly
shorter relative  to that given in Eq.~(\ref{Qtime}).  

The origin of quantum breaking time can be understood in general terms 
by viewing the semi-classical BHNP as a many-body system
 \cite{Dvali:2017eba}. 
In essence, it arises from  the internal rescattering of the constituents of  
the condensate which  leads to its depletion as well as to the generation of entanglement.  It also results in an effective 
``eigenstate thermalization"~\cite{Srednicki_1994,Murthy:2019fgs}.  Although the quantum state of the system is always pure, it evolves close to a thermal state in the sense that 
distinguishing it from a true thermal state requires precision  measurements 
over macroscopically long time scales. We will discuss these points further shortly. 

In the BHNP framework, the equivalence of Eq~ (\ref{time}) to 
Eq.~(\ref{Qtime}) can be understood as follows.  The rate of rescattering
of a pair of constituents can be estimated as 
\begin{equation} \label{gamma} 
\Gamma \sim Q_S \,\alpha_{\rm gr}^2\,N^2 \sim R_S^{-1} \,,
\end{equation} 
where the characteristic momentum transfer in our case is $Q_S$ and the Bose 
enhancement factor $\sim N^2$ comes from the number of pairs. Taking into account Eq.~(\ref{universal}), 
  during the time 
 $t_Q \sim N \Gamma^{-1} \sim N R_S$, an $O(1)$
fraction of constituents rescatters and the condensate half depletes. 
This gives Eq.~(\ref{Qtime}) and simultaneously establishes its equivalence to Eq.~(\ref{time}). 
Most of the depleted gravitons leave the condensate and escape in the form of Hawking quanta. With each scattering and emission, the remaining constituents become 
more and more entangled amongst each other. The entanglement reaches a maximal value after most of the gravitons have experienced 
collisions, with time scale given by Eq.~(\ref{time}).  Beyond this point, the semi-classical description of the 
system breaks down completely and  it ``quantum breaks". 

\subsection{Quantum breaking of Glasma formed in collisions of CGCs}

In the QCD case, the thermalization process is best studied in the ultrarelativistic collisions of large nuclei, which can be represented as 
colliding CGC shockwaves\footnote{One can also formulate the issues here for hadron-hadron scattering at sufficiently high energies where $R_S \ll R$, the hadron radius. While in 
the asymptotic Regge limit this system should thermalize to a Quark-Gluon Plasma, whether it does so in high energy proton-proton collisions is still open because its not clear that the required parametric separation of time scales is cleanly achieved even at the highest LHC energies.}. In these collisions,  a macroscopic deconfined region of high occupancy gluon fields  called the Glasma~\cite{Lappi:2006fp,Gelis:2006dv} is formed on a 
time scale $\tau\sim 1/Q_S$ corresponding to the longitudinal size of the wee gluon modes in the shockwaves. How this Glasma evolves and the thermalization process of 
the system into a Quark-Gluon Plasma (QGP) is an outstanding problem~\cite{Berges:2020fwq}. 

In the CGC picture, the collision of the shockwaves, and the subsequent evolution of the Glasma, can be studied numerically in a classical-statistical approximation at early times when the occupancies $N\gg 1$~\cite{Krasnitz:1998ns}. As we will discuss further later, Weibel-like plasma instabilities~\cite{Romatschke:2005pm} play a significant role in the decoherence of the Glasma, which then subsequently evolves, just as in the BHNP, through the rescattering of its constituents. Detailed 3+1-D numerical simulations~\cite{Berges:2013eia,Berges:2013fga} of the Glasma strongly suggest that this scattering occurs 
through a ``bottom-up" thermalization scenario~\cite{Baier:2000sb}  which we will now outline below. 

We shall closely follow 
\cite{Baier:2000sb} with a slight change to their notation, with $N_h$ of \cite{Baier:2000sb} denoting here the  occupation number instead of the number density, with the latter denoted 
instead by the lowercase $n_h$. The evolution in \cite{Baier:2000sb} starts from an initial saturated state 
of gluons that are referred to as ``hard", corresponding to momenta $p\sim Q_S$.   
Their number density in the initial Glasma stage decreases due to two effects. The first is the one-dimensional 
``classical" expansion\footnote{This is a consequence of the geometry of the collision which leads to a ``boost-invariant" expansion~\cite{Bjorken:1982qr}.} of the system which changes the number density 
as, 
\begin{equation} \label{Glasma:expansion} 
  n_h \sim \frac{Q_S^3}{\alpha_S} \frac{1}{(Q_St)}  \,,
\end{equation} 
where $t$ must be understood as the proper time. (In the absence of interactions, the occupation number is of course  conserved and is given by $N_h = \frac{n_h}{Q_S^2p_z} = 
\frac{1}{\alpha_S}$.)  The other effect influencing the decrease in the number density is the quantum scattering of gluons. 

 The only difference in the evolution of the Glasma to that 
of a Black Hole is the 
absence of the 1-D expansion. As explained above, the sole mechanism for a Black Hole to  lose its quantum constituents is through the depletion of the graviton 
condensate by their rescattering. The gravitons that leave the condensate in this way are responsible for Hawking radiation.  
The depletion of the graviton condensate due to rescattering  is strikingly similar to the initial time evolution of the CGC/Glasma,  which we shall now review. Of course, unlike Hawking quanta, the  rescattered gluons cannot escape to infinity unscathed, but form a QGP first. This is an inessential difference, as we shall discuss, from the ``early-time" perspective of our UV/IR correspondence. The rest of the early-stage dynamics in the two systems are qualitatively identical. 
 
When the occupancies are large, the net effect of number changing processes is small, and the scattering of the hard gluons is described by $2\leftrightarrow 2$ kinetic processes with the lowest momentum exchanged given by the Debye mass 
\begin{equation} \label{DB} 
  m_D^2  \sim \frac{\alpha_Sn_h}{Q_S} \sim  \frac{Q_S^2}{(Q_St)}  \,.
\end{equation} 
The frequency of collisions, per unit volume, is
  \begin{equation} \label{Coll} 
 \frac{d{\mathcal N}_{\rm col}}{dt} 
    \sim  \sigma n_h N_h \sim \frac{\alpha_Sn_h}{m_D^2p_z t}  \,,
\end{equation} 
where $\sigma \sim \alpha_S^2 m_D^{-2}$ is the scattering cross-section and the occupancy 
$N_h = \frac{n_h}{Q_S^2p_z}$.
 
Since $m_{\rm D}$ rapidly becomes smaller than $Q_S$, the net effect of the scatterings is to increase the longitudinal momentum of gluons (which otherwise would be 
rapidly depleted at the rate $p_z\sim 1/t$) via the random walk relation, 
\begin{equation} \label{csection}
 p_z^2  \sim  {\mathcal N}_{\rm col} \,m_D^{2} \sim \frac{\alpha_S n_h}
 {p_z}\,, 
\end{equation} 
and therefore, 
\begin{equation} \label{csection}
 p_z \sim  (\alpha_S n_h)^{\frac{1}{3}} 
 \sim \frac{Q_S}{(Q_St)^{\frac{1}{3}}}\,.  
\end{equation} 
Thus the  occupation number of hard gluons diminishes in time as 
\begin{equation} \label{total}
 N_h \sim  (\alpha_S)^{-1}(Q_St)^{-\frac{2}{3}}\,.  
\end{equation} 
The semi-classical picture of the Glasma as a high occupancy state with $N_h \sim \frac{1}{\alpha_S}$ is no longer tenable 
when 
$N_h\sim 1$, which occurs when  
\begin{equation} \label{timeCGC}
t_{\rm Glasma} \sim Q_S^{-1} \alpha_S^{-{3\over 2}}\, .
\end{equation} 
This time-scale is the ``quantum breaking" time of the CGC.

This quantum breaking time in the Glasma should however be considered an upper bound, because there is an overlap between the classical and 
quantum descriptions~\cite{Mueller:2002gd,Jeon:2004dh} that begins parametrically sooner as long as the occupancy satisfies $1\ll N_h \ll 1/\alpha_S$. 
As a lower bound, we can ask what the time scale is if rescattering of soft quantum is the only mechanism driving the depletion of the Glasma, as in the case of 
a Black Hole. This time-scale is
\begin{equation} \label{timeCGC}
t_{\rm Glasma}^{\rm low} \sim N/\Gamma =
 \frac{N}{(N^2\alpha_S^2) Q_S} =  \frac{1}{Q_S \alpha_S}\, , 
\end{equation} 
which is not surprisingly the same as that in Eq.~(\ref{Qtime}). One thus expects quantum breaking to occur in the Glasma in the temporal window $\alpha_S^{-1} < Q_S\,t < \alpha_S^{-3/2}$. 

The subsequent evolution and bottom-up thermalization of the Glasma into a Quark-Gluon Plasma is described by quantum kinetic $2\rightarrow 3$ processes captured by an effective 
kinetic theory~\cite{Arnold:2000dr}. A parametric estimate~\cite{Baier:2000sb} of the thermalization time gives $t_{\rm thermal}\sim \alpha_S^{-13/5} \,Q_S^{-1}$. Since $\alpha_S$ only grows logarithmically, no power of the coupling can beat the $Q_S^{-1}$ term, so asymptotically, in the Regge limit, thermalization in heavy-ion collisions is extremely rapid relative to the lifetime of the system $t_{\rm QGP}\sim R_A$, the nuclear radius. This ``late-stage" dynamics is of course qualitatively different from that of a Black Hole. 

\subsection{Connecting the time scales}
We shall now construct a dictionary between these  time scales in the two systems (CGC/Glasma versus BHNP).  To do so, we will describe the analogous process in the graviton condensate picture of a Black Hole. 
 There are two inessential differences. The first is that no decrease in the occupation number analogous to that in Eq.~(\ref{Glasma:expansion}) takes place  due to the absence 
of the classical expansion of the system\footnote{Fundamentally, this is because all the scattered partons in the $2\rightarrow N$ high energy graviton scattering are ``wee" partons; in QCD, a substantial fraction of the momentum is carried by hard ``valence" partons.}.  So for gravitons,  quantum rescattering is the only source for reducing their occupancy.  

 The other difference, of no great import to our dictionary, is that due to the isotropy  
 of the system, all relevant momenta of constituent gravitons  are always of order $R_S^{-1}$. As we discussed previously, this is also true in the CGC/Glasma for $t\leq 1/Q_S$, but 
 $p_z$ depletes rapidly subsequently due to the 1-D expansion. In order  to translate  the CGC results  into analogous results  for graviton condensate, we must take 
 \begin{equation} \label{relation} 
 p_z \sim Q_S \sim R_S^{-1}\,.
 \end{equation}
   
Next, in order to make contact with Eq.~(\ref{DB}), we  need to evaluate the equivalent  
mass of the many-body graviton state analogous to the QCD Debye scale. In this context, the origin of this mass is a spontaneous breaking 
of space-time translation symmetries by the graviton condensate. 
 As a result of this breaking, a graviton propagating through the condensate 
 mixes with the corresponding Goldstone mode and becomes 
 effectively massive. However, unlike the ordinary Higgs effect, the 
 Goldstone is not an external field but comes from the collective excitations 
 of the graviton condensate itself\footnote{This is analogous to the ``polaron" concept in condensed matter physics, with the broken translational symmetry reflected in the 
 localized polaron wavefunction.}. 
 
 The strength of the effect can be estimated 
 from the coupling of the graviton $h_{\mu\nu}$  with its own energy momentum tensor. It is sufficient to evaluate the latter quantity 
  $T^{\mu\nu} = \partial^{\mu} h_{\alpha\beta}\partial^{\nu}h^{\alpha\beta} \, 
 +...$, to bilinear order. The details of the tensorial structure are not important for our purposes
 and Lorentz indices will be suppressed henceforth.   It is sufficient to remember that all the interaction terms contain only two derivatives. The order parameter of spontaneous breaking of translation symmetry is represented by the 
 expectation value of the derivative of a canonically normalized graviton field over  the condensate state $\langle \partial h \rangle \sim \sqrt{ 
 \frac{n}{R_S}}  \sim \frac{\sqrt{N}}{R_S^2}$. As a result, the various components of the graviton field derivatively mix with each other and this generates a Debye screening effect analogous to  a thermal 
 plasma, 
 \begin{equation} \label{hmix}
   {1 \over M_P} h_{\mu\nu}  T^{\mu\nu} = 
      {\sqrt{n} \over M_P \sqrt{R_S}} h\partial h  \, +... = m_D h\partial h  \, + ...\,. 
 \end{equation} 
Remembering that $\alpha_{\rm gr} = (R_SM_P)^{-2}$
and taking the number density at the saturation point of  critical packing, $n = \frac{N}{R_S^3} = \frac{1}{\alpha_{\rm gr}R_S^3}$,
we can write, 
\begin{equation} \label{DBg}
   m_D^2 = \alpha_{\rm gr}  n R_S \, =  R_S^{-2} = Q_S^2\,.
 \end{equation} 
We see therefore an exact analogy between 
the Debye mass and the first relation in both Eqs.~(\ref{DB}) and (\ref{DBg}), and the equivalence of the second relation in the former to the latter at $t= 1/Q_S$, which is the minimum 
crossing time of two CGC shockwaves.
 
 It is easy to see that the expressions for scattering rates in the two systems also match. Indeed,  the scattering cross-section of constituent 
 gravitons is
 \begin{equation} \label{csection}
 \sigma \sim \alpha_{\rm gr}^2 R_S^{2} \sim \alpha_{\rm gr}^2  m_D^{-2} \,, 
\end{equation} 
 where in the last equation we took Eq.~(\ref{DBg}) into account.  
 The resulting interaction rate
 per unit volume is, 
  \begin{equation} \label{Collgr} 
 \frac{d{\mathcal N}_{\rm col}}{dt} 
    \sim  \sigma n N  \sim \frac{\alpha_{\rm gr} n}{m_D^2}  \,,
\end{equation} 
which by Eq.~(\ref{relation}) is the same as 
Eq.~(\ref{Coll}). 

Taking into account critical packing, it is not hard to see that the above expression fully matches the depletion rate given by Eq.~(\ref{gamma}), which is the source of 
 Hawking evaporation.  The corresponding half-depletion time is therefore given by Eq.~(\ref{time}). 
 The reason why the analogous time in Eq.~(\ref{timeCGC}) in the Glasma is faster by a factor
 $\sim \alpha_S^{1/3}$ is due to its one-dimensional expansion, which contributes to its dilution; as we argued previously, this factor should be considered an upper bound. 

\subsection{Lyapunov exponents, quantum breaking, chaos and scrambling} 

We  now turn to parallels between the two systems regarding chaos and information scrambling.  
The concept of Black Holes as information scramblers was introduced by  Hayden and Preskill \cite{Hayden:2007cs} who
suggested that the thermalization time of perturbed Black Holes
is bounded from below by 
 \begin{equation} \label{stime}
 t_{\rm sc}^{\rm BH}  =   R_S \ln(S_{BH}) \,,
 \end{equation} 
where $R_S$ is the gravitational radius and $S_{BH}$ is its entropy. 
 Later,  Sekino and Susskind \cite{Sekino:2008he} suggested that Black Holes 
 actually saturate this bound. 
  However, the underlying quantum mechanism behind Black Hole's fast scrambling  
  remained unclear for some time due to the lack of a suitable microscopic theory of a Black Hole.  The formulation of the Black Hole quantum $N$-portrait made it possible
  to establish such a connection. 
  This was done in \cite{Dvali:2013vxa},  where 
 the physical meaning of fast scrambling was identified with 
 the  {\it quantum break-time} and chaos. 
 That is, scrambling has been linked with a significant 
departure of the quantum evolution from  
the classical one. It was proposed in this work that the Lyapunov exponent
 $\lambda$ must play the crucial role in making the system
a fast scrambler.  The scrambling time can therefore be expressed as \cite{Dvali:2013vxa}, 
    \begin{equation} \label{Ltime}
 t_{\rm sc}^{\rm BH}  =   \lambda^{-1} \ln(S_{BH}) =  
 \lambda^{-1} \ln(N) \,.
 \end{equation} 
The very last equality uses the fact that in $N$-portrait 
the Black Hole entropy  
is equal to the number of its constituent gravitons $S_{BH} = N$   
 and reflects the fact that the scrambling time can be represented solely in terms of 
 a Lyapunov exponent and the number of constituents of the system, 
 without reference to any of its other characteristics.  
 Some related ideas about Lyapunov exponent were further 
 suggested in \cite{Maldacena:2015waa,Shenker:2014cwa,Stanford:2015owe}. 
    
  As a result, the formula  
  (\ref{Ltime}) for the scrambling time can be 
  generalized to arbitrary systems beyond Black Holes. 
  In particular, this equation tells us that any  
  critically packed condensate with occupation number $N$ 
  and a Lyapunov exponent $\lambda$  is a fast scrambler.  
  In  \cite{Dvali:2013vxa} this was explicitly demonstrated on a 
 $1+1$ dimensional Bose-Einstein condensate that 
 was originally introduced in \cite{Dvali:2012en} 
 as a prototype toy model for a Black Hole portrait.  It was discovered
 that an initial classical state of 
 an uniform condensate would develop a Lyapunov exponent 
 for the regime $N\alpha > 0$. Then, it would   
 undergo quantum breaking and chaos and would reach the state of maximal entanglement, after the time 
 given by Eq.~(\ref{Ltime}).  
 
  When applying the above knowledge to a black hole, 
  we must keep in mind that there may exist several 
  Lyapunov exponents that can 
  have potentially different physical origins, depending on the situation 
  in which the black hole is placed.  For example, in case of a
  classically-perturbed black hole, a natural candidate(s) for 
  $\lambda$ is the inverse dissipation time(s) for a quasi-normal 
  mode(s).  As an extreme form of the disturbance we can think of a black hole merger.   Correspondingly, we have to be careful 
  when drawing analogies with the similar time-scales 
  in the CGC/Glasma to which we shall turn now. 
  As we shall see, the appearance of a $\ln(N)$ factor in these time-scales 
 is universal there.    
  
When the spacelike CGC fields generate the timelike Glasma fields in nuclear collisions, the initial state of the matter after $t\sim 1/Q_S$, because of its occupancy of $O(1/\alpha_S)$,
is that of a coherent classical field. This field develops subsequently to the bottom-up scattering and thermalization scenario through a process of information scrambling and entanglement. Because of its  initial 1-D geometry, with matter exploding undergoing a Hubble-like expansion into the vacuum, the $p_z\sim Q_S$ modes are rapidly redshifted relative to the transverse momentum modes with typical $p_T\sim Q_S$. This anisotropy triggers an instability whereby quantum fluctuations,  initially $\alpha_S$ suppressed relative to the coherent Glasma condensate, grow nearly exponentially as $\sim e^{\sqrt{Q_S t}}$~\cite{Romatschke:2005pm,Romatschke:2006nk}. These quantum fluctuations therefore become of the order of the classical field on the 
time scale $t_{\rm sc}^{\rm Glasma} \sim Q_S^{-1} \ln^2(\alpha_S^{-1})$ -- corresponding to a butterfly effect that qualitatively changes the physics. This differs from $t_{\rm sc}^{\rm BH}$ only by the additional logarithmic factor-a consequence of the expanding geometry of the former\footnote{In a fixed box geometry, these instabilities indeed grow exponentially with $t$~\cite{Mrowczynski:1993qm,Arnold:2003rq,Berges:2007re} as opposed to 
$\sqrt{t}$ in the expanding case.}. 

In the framework of classical statistical field theory for nonequilibrium fields~\cite{Berges:2004yj,Aarts:2001yn}, these exponentially growing quantum fluctuations can be resummed to all orders in $\alpha_S e^{\sqrt{Q_S t}}$ and expressed as a initial spectrum of fluctuations for the evolution of the classical fields~\cite{Dusling:2011rz,Epelbaum:2013waa}. Each time evolution of a classical field configuration $A_{\rm cl}(t)\sim 1/\sqrt{\alpha_S}$ is seeded with a small quantum fluctuation $a \sim O(1)$ at the initial time $t= 1/Q_S$ drawn from this initial spectrum of fluctuations. Averaging observables over the spectrum of fluctuations leads to the randomization of the information contained in the Glasma. This is seen very cleanly in a toy 
0+1-dimensional $\phi^4$ model where the field equations can be solved analytically and correspond to Jacobi elliptic functions; in this simple setup, the difference between two seeds can be mapped on  to a small initial phase difference in classical trajectories in the Poincar{\'{e}} phase plane of the phase space density. With time evolution, the phase difference between the trajectories grows rapidly, leading to decoherence of the trajectories and the filling up the Poincar{\'{e}} plane~\cite{Dusling:2011rz}. 

This phenomenon, dubbed  ``prethermalization"~\cite{Berges:2004ce}, causes such high occupancy fields to  flow to a non-thermal fixed point described by the quasi-stationary self-similar behavior of single particle distributions characteristic of weak wave turbulence~\cite{Zakharov}; for the 3+1-D Glasma, numerical simulations~\cite{Berges:2013eia,Berges:2013fga} show that the corresponding single particle distributions are precisely those that describe the early-stage of bottom-up thermalization we discussed earlier. Remarkably, this 
non-thermal fixed point appears universal, and is seen to be identical to that of overoccupied self-interacting scalar fields~\cite{Berges:2014bba}. The subsequent scattering generates further entanglement, whose entanglement entropy grows until the system thermalized completely. 

\section{Outlook}
In this work, we outlined some of the elements of a remarkable correspondence between the saturated wee gravitons that constitute the Black Hole Quantum $N$-Portrait~\cite{Dvali:2011aa, Dvali:2012en}
and saturated wee gluons that constitute a Color Glass Condensate state~\cite{Gelis:2010nm} 
 inside hadrons that can be probed at very high energies. Both systems can be understood as classical states generated in 
$2\rightarrow N$ scattering when $N= 1/\alpha$, which also corresponds to the unitarization boundary for the scattering. A key feature of the correspondence is an emergent saturation scale $Q_S$ that characterizes the universal dynamics of the two systems on the unitarization boundary. 

The universal feature shared between the two systems is 
the saturation of the unitarity bounds  Eq.~(\ref{USB}) and
Eq.~(\ref{Gf})  on micro-state entropy
and the attainment of the relation Eq.~(\ref{universal}) for a generic saturated system \cite{Dvali:2020wqi}, \cite{Dvali:2019jjw, Dvali:2019ulr}.  
In particular, this implies that both objects
exhibit the area form of the entropy in terms of 
respective Goldstone constants $f$ of spontaneously broken 
Poincar{\'{e}} invariance.  In the case of the BHNP, 
where the Goldstone boson comes from graviton with the decay constant  $M_P$, 
Eq.~(\ref{Gf}) reproduces the Bekenstein-Hawking entropy. 
Analogously, the area law of entropy in the CGC is 
determined by the Goldstone scale that controls the  gluon screening 
length.  

 The minimal time-scales required for retrieving quantum information from the two systems saturate the general bounds 
  in Eq.~(\ref{tG}) and Eq.~(\ref{tAl}) derived in \cite{Dvali:2020wqi}. 
 As already noticed there for a Black Hole, both expressions reproduce 
 Page time \cite{Page:1976df}.   Further, the decay of these semi-classical states through scrambling and scattering  also is strongly analogous (modulo differences due to the different geometries of the two systems) until a quantum breaking time, beyond which time the two systems differ significantly. 

We showed that the observed universality provides deep insight into the complementary languages used to describe the two systems and their mapping into each other. For instance, the universality enabled us to conjecture that the entropy of the CGC saturates the unitarity bounds 
Eq.~(\ref{USB}) and Eq.~(\ref{Gf}) and allowed us to interpret the onset of screening and shadowing in this system in terms of a Goldstone scale for the breaking of Poincar{\'{e}} invariance; this establishes the area law for the CGC entropy in a given rapidity interval. 

The CGC-BHNP correspondence can also provide a quantitative understanding of unitarization in the BHNP, employing key features of a double copy between QCD and gravity amplitudes. In particular, a classical double copy between the two theories, may allow one to demonstrate, employing the dynamics of saturation in QCD, the onset of classicalization and unitarization in gravity amplitudes. This work is in progress and will be reported separately. 

An interesting question is whether our work sheds light on a suggestion in  \cite{Alvarez-Gaume:2006klm,Alvarez-Gaume:2008emf}
  that there is an equality  between the critical exponents, as observed by Choptuik \cite{Choptuik:1992jv}, appearing  in the self-similar  collapse of classical fluids in $D > 4$ dimensions and the exponent governing the growth of the BFKL equation in $D=4$ dimensions.  The principal point of our work  however is that the 
CGC and Black Holes in BHNP are structurally similar 
states at the full quantum level in $D=4$ dimensions.  This follows from the universal behavior of these objects saturating the entropy bounds in 
Eq.~(\ref{USB}) and Eq.~(\ref{Gf}). The saturation of the entropy bounds implies the unitarization of the corresponding 
$2 \rightarrow N$ processes (gluon and graviton) in weak coupling in both QCD and gravity in $D=4$ dimensions. 
It would be interesting to explore whether it is at all possible that the saturation of the entropy and unitarity bounds translates into values of classical critical exponents.

A striking connection between the BHNP and the CGC that deserves to be explored further is that between gravitational memory and color memory, and their connection to soft theorems governing the infrared structure of scattering amplitudes. Strominger has argued that an infrared ``triangle" relates the asymptotic BMS symmetries of gravity to the soft theorems and the gravitational memory effect, and conjectured that this infrared triangle is a universal feature of gauge theories~\cite{Strominger:2017zoo}. The connection of this picture to the BHNP was discussed in \cite{Averin:2016ybl}. As argued there, the Goldstone modes on the Black Hole event horizon can be identified with those of spontaneously broken {\it horizon} supertranslations. 

The mapping of gravitational memory to color memory was first discussed in \cite{Pate:2017vwa} in the context of Yang-Mills fields on the celestial sphere at null infinity. 
It was shown in \cite{Ball:2018prg} that the classical equations describing the transition between degenerate vacua in the infinite momentum frame are precisely the equations describing the CGC classical field.  Further, just as gravitational memory can be measured as a physical displacement of inertial gravitational wave detectors, the color memory of the CGC state 
is measurable  due to the large transverse momentum kick $p_\perp \sim Q_S$ delivered to a colored probe such as a quark-antiquark pair that experiences the passage of a CGC shock wave. The relation of color memory in the CGC to the color memory of gauge fields at null infinity has been argued on the basis of the CGC fields being static in lightcone time. However the mapping of the CGC to the BHNP suggests that there is a further interpretation of information carrying CGC modes as Goldstone modes of spontaneously broken Poincar{\'{e}} symmetry. This feature of the BHNP-CGC correspondence deserves further investigation. 

\section*{Acknowledgements}

This work of GD is supported in part by the Humboldt Foundation under Humboldt Professorship Award, by the Deutsche Forschungsgemeinschaft (DFG, German Research Foundation) under Germany's Excellence Strategy - EXC-2111 - 390814868,
and Germany's Excellence Strategy  under Excellence Cluster Origins.
 
RV is supported by the U.S. Department of Energy, Office of Science, Office of Nuclear Physics, under contract No. DE- SC0012704. His work was also supported by the Humboldt Foundation through a Humboldt Prize and by the DFG Collaborative Research Centre SFB 1225 (ISOQUANT) at Heidelberg University. RV would like to thank Alan Caldwell, the Max Planck Institute and Ludwig-Maximilians University in Munich for their kind hospitality. He would also like to thank Himanshu Raj, Vladi Skokov and Kong Tu for valuable discussions. 

\section{Appendix A. Goldstone modes and the area law entropy bound}

We will briefly review here the arguments of 
 \cite{Dvali:2020wqi} about the emergence of Goldstone modes and their role in the area-law form Eq.~(\ref{Gf}) of the entropy bound  
 Eq.~(\ref{USB})  
 for a generic critically packed system Eq.~(\ref{NA}), thereby 
supporting the relation Eq.~(\ref{universal}).

 Consider a bosonic field $\phi_j$
described by creation/annihilation operators $\hat{a}_j(\vec{k})^\dagger, 
\hat{a}_j(\vec{k})$, 
 \begin{equation} \label{expansion}
 \hat{\phi}_j  = \sum_{\vec{k}} {1\over \sqrt{2V\omega_{\vec{k}}}}
 \left ( {\rm e}^{i\vec{k}\vec{x}} \hat{a}_j(\vec k)  +  {\rm e}^{-i\vec{k}\vec{x}} \hat{a}_j(\vec k)^{\dagger} \right ) \,.  
 \end{equation} 
 Here $V$ is the volume and the label $\vec{k}$ refers to the three-momentum. The  label $j=1,...,n$ accounts for different 
spin-polarizations and for a representation content 
with respect to internal symmetries such as color 
or flavor.  The modes obey the standard 
bosonic commutation relations, 
   $ [\hat{a}_i(\vec{k}),\hat{a}_j(\vec{k'})^{\dagger}] = \delta_{ij}\delta_{
    \vec{k}\vec{k'}} \,,
  [\hat{a}_i(\vec{k}),\hat{a}_j(\vec{k'})]  = 0 $. 
 We shall assume that the field interacts via a coupling 
 $\alpha$.  
 
 We shall use states in which modes of certain momenta $\vec{k}$  are macroscopically occupied, 
 \begin{equation}\label{nstate}
 \ket{N} = 
 \prod_{j=1}^n {(\hat{a}_j(\vec{k})^{\dagger})^{N_j} \over \sqrt{N_j!}} \ket{0}\, , 
\end{equation}
where $N$ refers to the  total occupation number, 
\begin{equation}\label{totaln}
N = \sum_{j=1}^{n} N_j   \,.
\end{equation}
Analogously, we can form a  coherent state  that 
   describes a classical field-configuration localized within a certain characteristic radius 
  $R$, 
  \begin{equation}\label{sol}
 \ket{sol} = {\rm e}^{\sum_{\vec{k}}\sum_{j=1}^n \sqrt{N_j(\vec{k})} (\hat{a}_j(\vec{k})^{\dagger}
 - \hat{a}(\vec{k})_j)} \ket{0} \,,
\end{equation}
with 
\begin{equation}\label{totalnk}
\sum_{j=1}^{n} \sum_{\vec{k}} N_j(\vec{k}) \, =\,  N \gg 1 \,,
\end{equation}
 where the $N_j(\vec{k})$'s are sharply peaked around the momentum $|\vec{k}| = \frac{1}{R} = Q_S$. \\ 
 
 The classical field is described by the expectation value
 over this coherent state, 
 \begin{equation} \label{VEV}
 \phi_j =  \bra{sol} \hat{\phi}_j \ket{sol}\,.
\end{equation}  
We shall refer to such a state as a
{\it soliton} state.  

 We shall assume that the occupation number  
 $N$ does not exceed $1/\alpha$. In this case, the kinetic energy of a given quantum, 
$E_{kin} \sim \frac{1}{R}$, is not subdominant  
to the potential energy of its interaction with the rest,  given by $E_{\rm pot} \sim \frac{\alpha N}{R}$.  
Under such conditions, the localized classical field configuration
$\phi_j$  represents a $N$-particle state of 
characteristic momenta $\sim Q_S = 1/R_S$, each contributing 
$\sim Q_S$ towards the energy of the soliton. Therefore the  total 
energy is,
\begin{equation} \label{energyn}
  E \sim {N \over R_S}  \sim  N Q_S \,.   
\end{equation}
Now according to \cite{Dvali:2020wqi}, unitarity puts an upper bound 
on entropy given by Eq.~(\ref{USB}). 
 This also agrees with the argument in \cite{Dvali:2019jjw}
that the entropy of a self-sustained $N$-particle state 
with $N \lesssim 1/\alpha$ is bounded by $N$. 

Let us now show that taking into account Eq.~(\ref{NA}), 
the entropy bound 
in Eq.~(\ref{USB}) is equal to the area in units of a Goldstone decay constant $f$. For this, we  must determine $f$. The localized classical field configuration $\phi_j$ always  spontaneously breaks Poincar{\'{e}} symmetry (and internal symmetries), with the strength of  the breaking of Poincar{\'{e}} invariance 
 measured by the gradient of the field 
 $\sum_j (\nabla \phi_j)^2 \sim \frac{N}{R^4 }$.  The corresponding Goldstone fields 
are collective modes emerging from the infinitesimal 
coordinate-dependent transformations.  The  scale $f$ is the coefficient 
accompanying the dimensionless parameter 
of such transformations. 

For example, consider translations along a 
coordinate $x$. For simplicity, in this case, 
we can drop the internal index $j$.
 The relevant infinitesimal variation of the solution is, 
 \begin{equation}\label{SHIFTx}
  \delta \phi   =   \epsilon_x R\partial_x\phi  = \epsilon_x f
  \end{equation}
  where $\epsilon_x$ is dimensionless.  Its coefficient 
 $f = R\partial_x\phi_j$ represents the decay constant of the Goldstone 
 field.  It is clear that   
  \begin{equation}\label{fV}
  f = R \partial_x \phi  =   \frac{\sqrt{N}}{R} \,.
 \end{equation} 
 Now taking into account Eq.~(\ref{NA}), it is obvious that we can 
 rewrite Eq.~(\ref{USB}), in form of the area law Eq.~(\ref{Gf}), 
 \begin{equation} \label{Asol}
 S_{\rm max} = \frac{1}{\alpha}  =  N =  (Rf)^2 \,,    
\end{equation}
thereby reproducing  
Eq.~(\ref{universal}) in its entirety.

 Due to the universality of the Goldstone language and unitarity,   
the above relations emerge as a generic property of the saturated
systems, regardless of their underlying structure, as has been confirmed for several systems \cite{Dvali:2019jjw, Dvali:2019ulr, Dvali:2020wqi}.  

Now, for both the BHNP and the CGC, the Goldstone decay   
 constant is given by Eq.~(\ref{fV}).  For the BHNP, this scale is equal to the Planck mass 
 $f =  \sqrt{N} Q_S = \sqrt{N}/R_S = M_P$. 
 It is easy to check that in terms of parameters $N, f$ and 
 $\alpha$, the effective $3$-graviton vertex 
 in BHNP and  $3$-gluon vertex in CGC take similar forms. 
 We remind the reader that a generic three-graviton vertex is 
weighted by two derivatives ($\partial^2$) and a single power of $1/M_P$. For the saturated state of BHNP, these translate as
 $\partial^2 \rightarrow Q_S^2$ and 
 $1/M_P = 1/f = R_S/\sqrt{N}$ respectively. Thus we have, 
  \begin{equation}
 {\rm three\,\,graviton \,\,effective \,\,vertex} = 1/(R_S^2M_P) = 
 \sqrt{\alpha_{\rm gr}} \,Q_S = 
 Q_S^2/f \,.
 \end{equation}
 This matches the expression for the strength of the 3-gluon vertex for CGC fields, 
 \begin{equation}
 {\rm three\,\, gluon\,\, effective\,\, vertex} =
 \sqrt{\alpha_S} \,Q_S = Q_S/\sqrt{N} = Q_S^2/f\,.
 \end{equation}
 Thus in units of the Goldstone scale, the strength of the coupling in the two saturated systems is identical.

\bibliography{BH-CGC}

\end{document}